\documentclass[3p,times,procedia]{elsarticle}
\usepackage{nupha_ecrc}

\volume{00}

\firstpage{1}

\journalname{Nuclear Physics A}

\runauth{}

\jid{nupha}

\jnltitlelogo{Nuclear Physics A}

\usepackage{amssymb}

\usepackage{lineno}

\usepackage[figuresright]{rotating}

\begin{document}

\begin{frontmatter}



\dochead{XXVIIIth International Conference on Ultrarelativistic Nucleus-Nucleus Collisions\\ (Quark Matter 2019)}

\title{First Direct Observation of the Dead-Cone Effect}


\author{Nima Zardoshti \\ on  behalf of  the ALICE Collaboration}

\address{}

\begin{abstract}
We report the first direct measurement of the dead-cone effect at colliders, using iterative jet declustering techniques in pp collisions at $\sqrt{s} = 13$ TeV. The procedure exposes the splittings of $\rm{D^{0}}$ mesons in the jet shower, by iteratively declustering the angular ordered C/A tree. The splitting history of the $\rm{D^{0}}$ meson initiated jet is mapped onto the Lund Plane, where appropriate cuts can be made to suppress hadronisation effects. The reported variable is the splitting angle with respect to the $\rm{D^{0}}$ meson axis, which is updated after each splitting. Track-based jet finding, along with the low transverse momentum reach for charged tracks of the ALICE detector, allow for an accurate reconstruction of the splitting angle in the phase-space where the dead-cone effect is expected to be largest. The results are compared to those of inclusive jets, where the dead-cone effect is expected to be negligible.
\end{abstract}


\end{frontmatter}


\section{Introduction}

The dead-cone effect is a fundamental property of all radiative theories. It defines an angle relative to the emitter, within which emissions are parametrically suppressed. This angle is defined as $m_{\rm{q}}/E_{\rm{q}}$, where $m_{\rm{q}}$ and $E_{\rm{q}}$ are the rest mass and energy of the emitter, respectively~\cite{deadcone}. This carries consequences for all massive emitters in a given theory. In QCD these emitters are the quarks, which have a non-zero mass. At the lower end of the energies accessible at ALICE, the effects are significant for charm and beauty quarks. The presence of a dead-cone region results in a harder fragmentation of these heavy quarks in the vacuum, due to the reduced available phase-space for emissions, and has implications for their energy loss when they traverse the deconfined Quark-Gluon Plasma medium.
\\
\\
A direct observation of the dead-cone effect at colliders had not been possible till now, due to a number of challenging experimental factors. Primarily, the dead-cone region is susceptible to pollution by both the decay products of the heavy quarks (and their subsequent emissions) and hadronisation effects. Another experimental constraint was the accurate determination of the emission axis, as this must be updated after each subsequent emission. Previous attempts at an indirect measurement of the effect were made at DELPHI~\cite{delphi}, using generalised axes such as the event thrust or the jet axis. Recently, new techniques for uncovering the dead-cone effect using jet reclusetering algorithms have been proposed. One technique involves the statistical separation of radiation emitted from heavy quarks in a 4-body top quark decay~\cite{top}. Another method~\cite{method}, which this analysis is based on, allows for the measurement of the dead-cone effect for charm and beauty quarks by taking advantage of Lund Planes~\cite{lundplanes}.

\section{Jet Splittings}

The jet splitting tree can be accessed by reclustering track-based jets with a given jet reclustering algorithm which maintains angular ordering. In such a configuration, unwinding the reclustering process orders the splittings chronologically. Two-dimensional maps of the splitting kinematic parameters, known as Lund Planes, can be constructed and filled per splitting. These maps allow for a visualisation of the splitting phase-space. The relevant splitting kinematics for this report are $\theta$, $E_{\rm{rad}}$ and $k_{\rm{T}}$, which denote the splitting opening angle, energy of the mother particle and transverse scale of the subleading prong, respectively. Comparing the angular axis of Lund Planes for heavy flavour tagged and inclusive jets, can provide an experimental handle on the dead-cone effect. This would manifest as a suppression of splittings at small angles for heavy-flavour tagged jets compared to their inclusive counterparts.

\section{Analysis}
\subsection{$\rm{D^{0}}$-tagged Jets}
This analysis was performed on minimum bias pp data at $\sqrt{s}=13$ TeV, with events comprising of charged tracks, with transverse momentum of $p_{\rm{T}}^{\rm{ch}} \geq 0.15 $ GeV/$\it{c}$, constrained within the ALICE TPC acceptance (pseudorapidity of $|\eta| < 0.9$). The $\rm{D^{0}} \rightarrow K\pi$ (and charge conj.) hadronic channel, which has a branching ratio of $3.89\%$, was used to reconstruct the $\rm{D^{0}}$ candidates which were selected using topological and PID cuts. These $\rm{D^{0}}$candidates have a transverse momentum range of $2 \leq p_{\rm{T,D}} < 36$ GeV/$\it{c}$. Jet finding was performed independently for each $\rm{D^{0}}$ candidate in an event, as if it was the only one. Prior to jet finding, the $\rm{D^{0}}$ daughters were replaced by a vector comprising of the sum of their four momenta, which is labeled as the $\rm{D^{0}}$ candidate. This improves the measured jet energy scale in cases where the opening angle between the daughters is larger than the jet radius. The event was then clustered into anti-$k_{\rm{T}}$ jets~\cite{antikt} with $R=0.4$ and a jet transverse momentum range of $5 \leq p_{\rm{T}}^{\rm{ch,jet}} < 50$ GeV/$\it{c}$. The jet containing the fully reconstructed $\rm{D^{0}}$ candidate was then selected.
\\
\\
A side-band subtraction procedure was then performed in bins of $p_{\rm{T,D}}$. In each $p_{\rm{T,D}}$ bin, the invariant mass of the $\rm{D^{0}}$ candidates was fitted with a Gaussian function for the signal and an exponential function for the background, with initial parameters taken from a Monte-Carlo simulation. The region within $2\sigma$ of the Gaussian peak was taken as the signal region whilst the regions at $4-9\sigma$ on either side of the peak were taken as the side-band regions. The side-bands are far enough away from the peak that they are considered to be fully background dominated. Consequently, they were used to model the background underneath the peak in the signal region.
\\
\\
Once the $\rm{D^{0}}$-tagged jets in the signal and side-band regions had been identified, they were reclustered with the Cambridge-Aachen algorithm~\cite{ca}. The reclustering history was unwound, following the prong containing the fully reconstructed $\rm{D^{0}}$ at each step, and the splittings in the jet were accessed. For each $p_{\rm{T,D}}$ bin, signal and side-band Lund Planes with axes of $E_{\rm{rad}}$ and $\ln(1/\theta)$ were filled. In order to remove non-perturbative effects, only splittings passing a given minimum $k_{\rm{T}}$ cut were accepted. The side-band Lund Plane was then scaled by a quantity reflecting the difference between the areas of the side-band regions and that of the background under the signal peak. The scaled side-band Lund Plane was then subtracted from the signal Lund Plane to obtain a subtracted distribution. Before summing the subtracted Lund Plane across the $p_{\rm{T,D}}$ bins, they were each scaled by the efficiency (PYTHIA~\cite{pythia} derived) of finding $\rm{D^{0}}$ tagged jets using the given selection cuts in that $p_{\rm{T,D}}$ bin. Finally this led to the $\rm{D^{0}}$-tagged jet Lund Planes in data.

\subsection{Inclusive Jets}
Jet finding and reclustering of the inclusive jet sample was performed with the same parameters as in the $\rm{D^{0}}$-tagged case. It was observed, with percent-level accuracy, that the fully reconstructed $\rm{D^{0}}$ was always found in the hardest prong of each splitting. Therefore when following the splitting tree of inclusive jets, at each step the hardest daughter prong was followed. Additionally, in order to mimic the same interaction $Q^{2}$ as was selected by the $p_{\rm{T,D}} \geq$ 2 GeV/$\it{c}$ cut, we required a $p_{\rm{T}} \geq$ 2.8 GeV/$\it{c}$ cut on the leading track of the hardest prong of each splitting entering the Lund Planes. This value corresponds to the transverse mass scale of a 2 GeV/$\it{c}$ $\rm{D^{0}}$. 

\section{Results}

\begin{figure}[ht]
\includegraphics[width=0.55\textwidth]{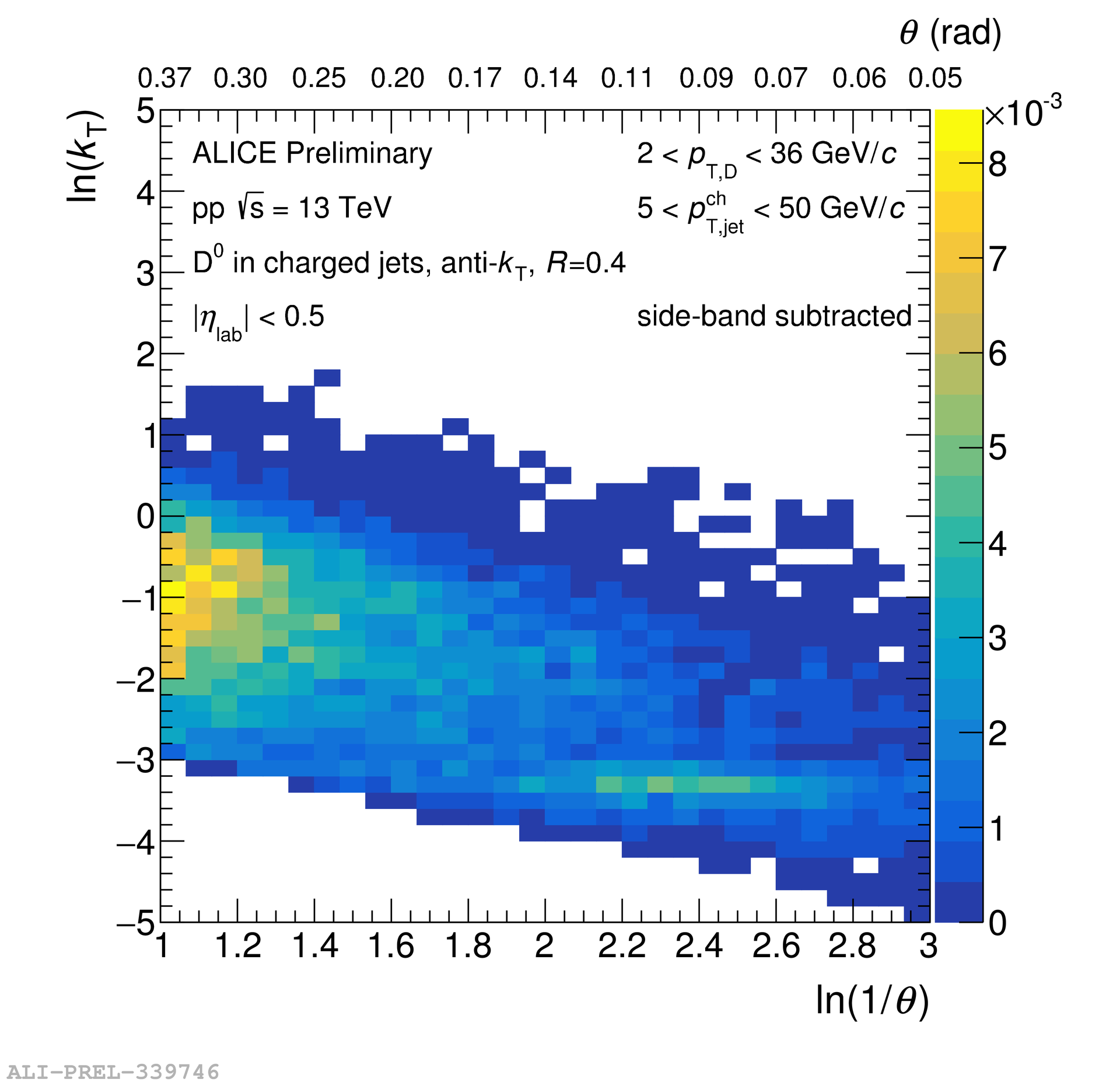}
\includegraphics[width=0.55\textwidth]{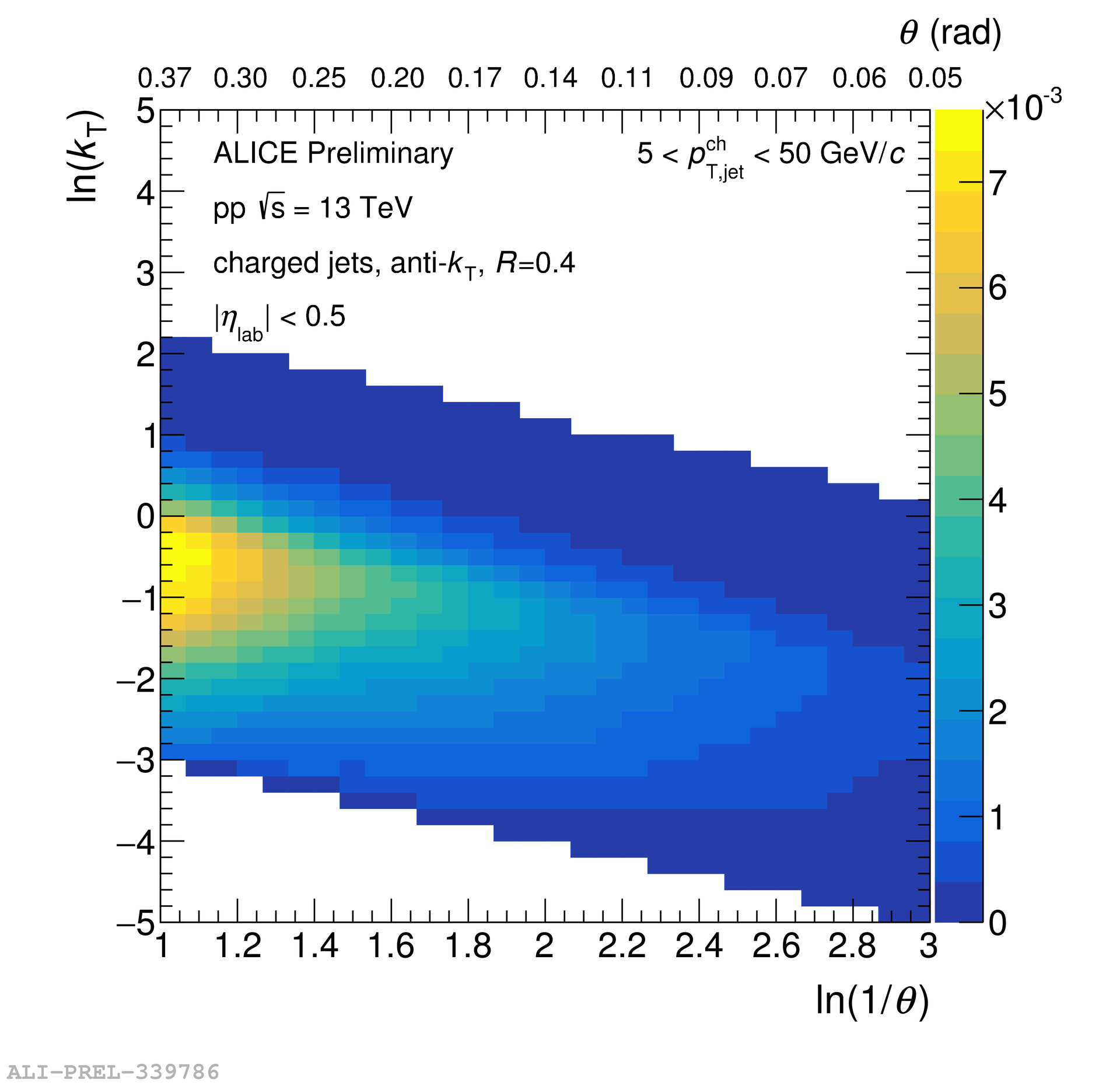}
\caption{Lund Planes for $\rm{D^{0}}$-tagged jets (left) and Inclusive jets (right) are shown.}
\label{fig:Lund_Maps}
\end{figure}

The final $\rm{D^{0}}$-tagged jet and inclusive jet Lund Planes are shown in Fig.~\ref{fig:Lund_Maps}. For illustrative purposes the Lund Planes shown here have $\ln(k_{T})$ instead of $E_{\rm{rad}}$ on the $y$-axis. At this stage already a visual hint of a suppression of splittings at small angles in the $\rm{D^{0}}$-tagged jets compared to inclusive ones can been observed. 
\\

\begin{figure}[ht]
\includegraphics[width=0.55\textwidth]{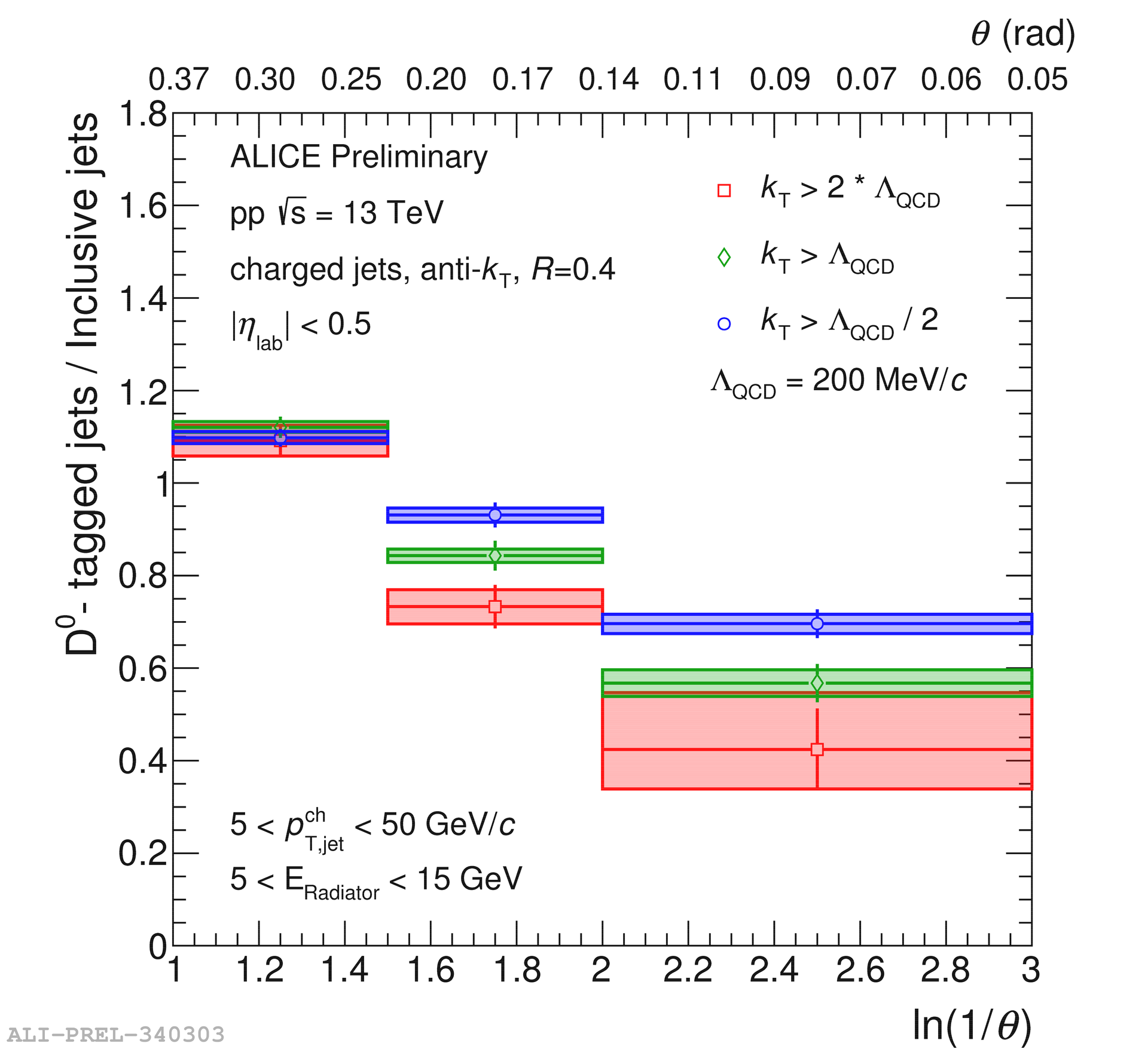}
\includegraphics[width=0.55\textwidth]{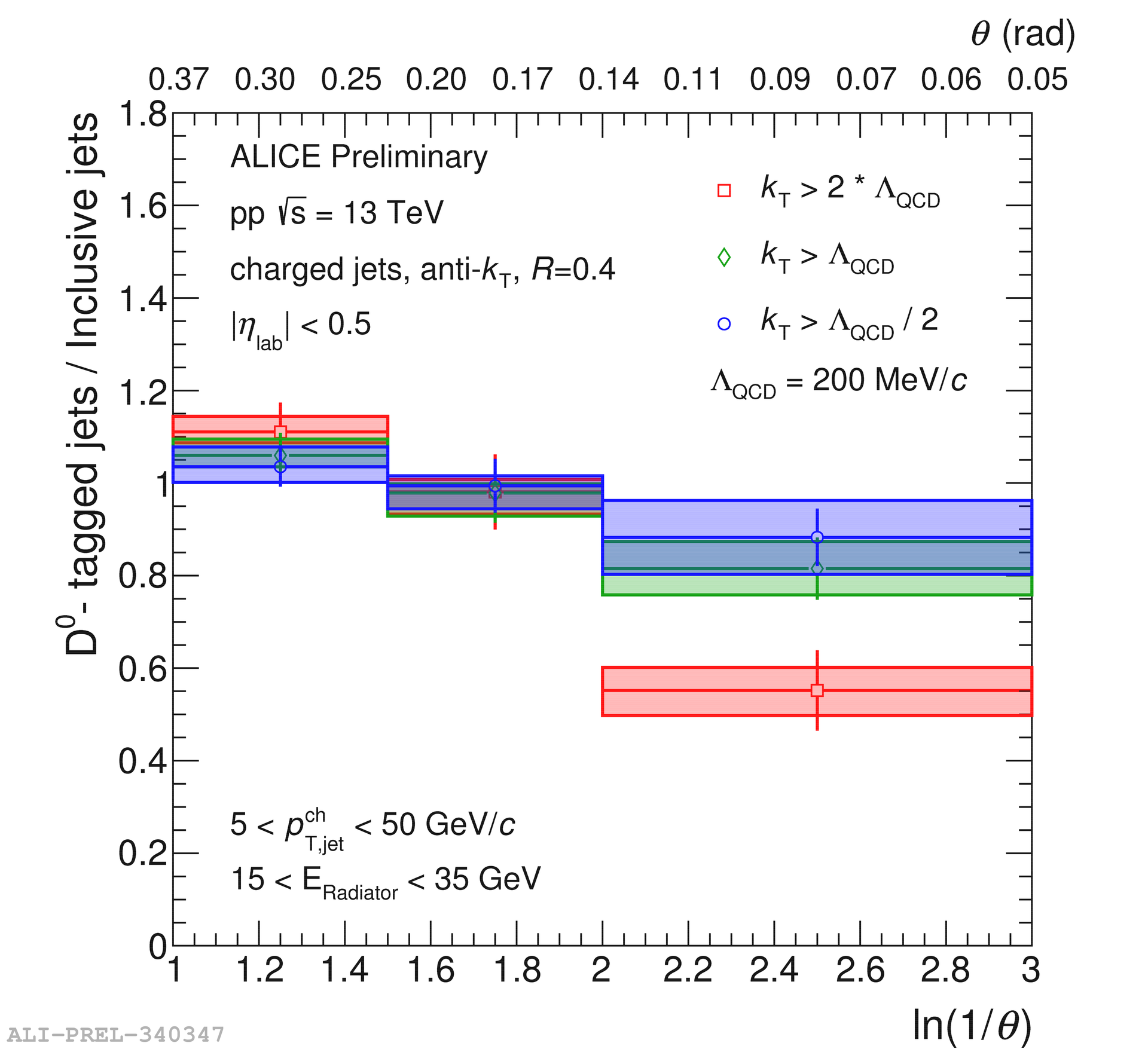}
\caption{Ratio of the angular distribution of splittings for $\rm{D^{0}}$-tagged jets vs inclusive jets are shown for $5 \leq E_{\rm{rad}} < 15$ GeV  (left) and $15 \leq E_{\rm{rad}} < 35$ GeV (right).}
\label{fig:DvsInclusive_Ratio}
\end{figure}

In order to directly observe the dead-cone effect, projections of the angular axis of the $\rm{D^{0}}$-tagged jet and inclusive jet Lund Planes were made in bins of $E_{\rm{rad}}$. Ratios of these projections were then taken, as shown in Fig.~\ref{fig:DvsInclusive_Ratio}. It can be seen that the distribution of splittings for the $\rm{D^{0}}$-tagged jets is suppressed at small angles compared to inclusive jets. As expected, this dead-cone effect is more significant for low $E_{\rm{rad}}$ splittings. Higher cuts of $k_{\rm{T}}$ remove more non-perturbative effects which in turn provide a cleaner dead-cone signal. However this comes at a cost to statistics.

\section{Conclusions}

We have reported the first direct measurement of the dead-cone effect by comparing the angular distribution of jet splittings involving a fully reconstructed $\rm{D^{0}}$ to that of inclusive jets. A strong suppression of splittings for $\rm{D^{0}}$-tagged jets at small angles is observed. The energy dependence of the dead-cone effect is also highlighted by comparing the suppression of splittings in different radiator energy bins.






 
\end{document}